\documentclass[preprint,showpacs,preprintnumbers,amsmath,amssymb,natbib]{revtex4}

\usepackage{graphicx}% Include figure files
\usepackage{epsfig}		
\usepackage{dcolumn}% Align table columns on decimal point
\usepackage{bm}% bold math
\usepackage{color}
\def\0{\mbox{\tiny $0$}}
\def\1{\mbox{\tiny $1$}}
\def\2{\mbox{\tiny $2$}}
\def\3{\mbox{\tiny $3$}}
\def\4{\mbox{\tiny $4$}}
\def\5{\mbox{\tiny $5$}}
\def\6{\mbox{\tiny $6$}}
\def\7{\mbox{\tiny $7$}}
\def\8{\mbox{\tiny $8$}}
\def\9{\mbox{\tiny $9$}}

\def\f14{\mbox{\tiny $\frac{1}{4}$}}

\begin{document}

\title{Equivalence between Born-Infeld tachyon and effective real scalar field theories for brane structures in warped geometry}

\author{A. E. Bernardini}
\email{alexeb@ufscar.br}
\altaffiliation[On leave of absence from]{~Departamento de F\'{\i}sica, Universidade Federal de S\~ao Carlos, PO Box 676, 13565-905, S\~ao Carlos, SP, Brasil.}
%\altaffiliation{Also at Instituto de F\'{\i}sica Gleb Wataghin, UNICAMP, PO Box 6165, 13083-970, Campinas, SP, Brasil}
\author{O. Bertolami}
\email{orfeu.bertolami@fc.up.pt}
\altaffiliation[Also at~]{Instituto de Plasmas e Fus\~ao Nuclear, Instituto Superior T\'ecnico, Av. Rovisco Pais, 1, 1049-001, Lisboa.} 
\affiliation{Departamento de F\'isica e Astronomia, Faculdade de Ci\^{e}ncias da
Universidade do Porto, Rua do Campo Alegre 687, 4169-007, Porto, Portugal.}
\date{\today}% It is always \today, today,

\begin{abstract}
An equivalence between Born-Infeld and effective real scalar field theories for brane structures is built in some specific warped space-time scenarios.
Once the equations of motion for tachyon fields related to the Born-Infeld action are written as first-order equations, a simple analytical connection with a particular class of real scalar field superpotentials can be found.
This equivalence leads to the conclusion that, for a certain class of superpotentials, both systems can support identical thick brane solutions as well as brane structures described through localized energy densities, $T_{00}(y)$, in the $5^{th}$ dimension, $y$.
Our results indicate that thick brane solutions realized by the Born-Infeld cosmology can be connected to real scalar field brane scenarios which can be used to effectively map the tachyon condensation mechanism.
\end{abstract}

\date{\today}
\maketitle

%\section{Introduction}

Brane cosmology driven by scalar fields has been recurrently studied in order to address the cosmological constant and hierarchy problems \cite{Arkani01,Arkani02}, as well as symmetry breaking issues \cite{Carvalho01} (see also Ref.~\cite{Carvalho02} for the projection on the brane of vector and tensor fields in the bulk space).
The first ideas for brane world scenarios assumed a warped $4$-dimensional brane universe embedded in a higher dimensional bulk space, where the brane corresponds to a localized delta function on the extra dimensional coordinate \cite{Randall}.
Brane world scenarios also have been discussed in the context of realizing $4$-dimensional gravity on a domain wall in $5$-dimensional space-time \cite{Randall,Cvetic01}, with extensions to domain walls in gravity coupled to scalars \cite{DeWolfe,Gold} and to time-evolving cosmological models \cite{Cosmo} (see also Ref.~\cite{Cvetic01} and references therein).

The brane scenario examined here is related to generic solutions of the $5$-dimensional Born-Infeld field theories of the form
\begin{equation}
S = \int dx^5\, \sqrt{\det{g_{AB}}}\,\left[-\frac{1}{4}R - U(\varphi)\sqrt{1 - g^{AB}\partial_A \varphi \partial_B \varphi}\right],
\label{000B}
\end{equation}
where $R$ is the scalar curvature, and $g_{AB}$ denotes the metric tensor, with $A$ and $B$ running from $0$ to $4$.
The field $\varphi$ is a tachyon field and $U(\varphi)$ is its potential, with dimensional constants absorbed by a suitable field normalization.
From this action, it has been conjectured that the dynamics of a Born-Infeld tachyon field in a background of an unstable $D$-brane system can be perturbatively described by the dynamics of an effective real scalar field \cite{Sen1}. 
According to such an assumption, tachyon calculations would be reliable only in the approximation where $\varphi$ derivatives can be truncated beyond the quadratic order \cite{Trun}.

The perturbative truncation leads to an effective action driven by a real scalar field, $\chi$, coupled to $5$-dimensional gravity, given by
\begin{equation}
S_{\tiny\mbox{eff}} = \int dx^5\, \sqrt{\det{g_{AB}}}\,\left[-\frac{1}{4}R + \frac{1}{2}g_{AB}\partial^A \chi\partial^B \chi - V(\chi)\right],
\label{000}
\end{equation}
which gives rise to several possibilities for investigating the related tachyon field dynamics.
In quantum field theories, a tachyon field can be realized by the instability of the quantum vacuum, described by the quantum state displaced from a local maximum of an effective potential like $V(\chi)$.
In the effective real scalar field scenario, the tachyon field would follow a spontaneous symmetry breaking (SSB) that implies into a process dubbed as tachyon condensation \cite{PRL1,PRL2}.
Given its remarkable applications in brane world models, tachyon condensation is argued to play an important role also in string theory (see e. g. Refs.~\cite{Witten,Polchinski}).
Tachyon condensation can also reproduce the results of a collision process similar to a kink-antikink or to a soliton-antisoliton annihilation that drives the system to the SSB vacuum after complete annihilation.
In this context, the Big-Bang has been hypothesized to be due to such a brane-antibrane collision.
Notice that branes defined as classical solutions of tachyonic potentials naturally arise in systems with rolling tachyons on unstable branes \cite{Gutperle}.
The resulting vacuum state after annihilation exhibits the remaining lower-dimensional branes as relics of tachyon condensation \cite{Sen} that (re)produce the effects of cosmic strings in brane cosmology \cite{Jones,Sarangi,Dvali}.

Real scalar field models coupled to gravity lead also to analytical solutions of gravitating defect structures which allow for the inclusion of thick branes used in several brane cosmology scenarios.
Thick domain walls, for instance, are often associated to integrable models.
In general, potentials associated to single real scalar field support BPS type solutions \cite{Bazeia,Brane} of first-order differential equations.
In this case, the equations result into topological defects that admit an internal structure.

However, there has been no consensus about how reliably effective real scalar field models can describe the Born-Infeld tachyonic dynamics \cite{Felder}, despite of the importance of real scalar fields in describing brane structures in warped geometry \cite{Hill,Djouadi,Gregory,DeWolfe,Gremm,Erlich,Campos}.

Therefore, the brane model discussed in this letter treats Born-Infeld tachyon fields without any build in association with the real scalar field (c. f. Eq.~(\ref{000})).
Assuming that the equations of motion for the Born-Infeld tachyon fields can be mapped by superpotential parameters constrained by first-order equations, analogously to the procedure of mapping BPS solutions into real scalar fields, one is able to find exact solutions for the tachyon field, $\varphi$.
In addition, a fruitful connection between tachyon and real scalar field superpotentials can be established.
The resulting brane scenario exhibits an exact equivalence between Born-Infeld tachyon and real scalar field dynamics in $5$-dimensions, which is reproduced by a unique warp-factor and leads to the same localized energy densities.

In what follows we shall call $\chi$ a real scalar field, even when considering that its associated action may approach a tachyonic action that circumstantially results into a condensation mechanism and associated instabilities.
We shall bear in mind that we seek for an analytical correspondence between the Born-Infeld tachyon with the real scalar field in order to obtain two equivalent brane world scenarios.

The framework for discussing a single real scalar field coupled to gravity in the brane scenario follows previous discussions \cite{Gregory,DeWolfe,Gremm,Erlich,Campos}.
The correspondence between the Born-Infeld tachyon and the real scalar field is obtained through a set of first-order equations.
Novel integrable models that admit thick brane solutions to the Born-Infeld action through twin warp factors bound from above are also discussed.

\subsection*{Real scalar fields}

Let us start considering a $5$-dimensional space-time warped in $4$-dimensions.
In order to ensure the Poincar\'e invariance in $4$-dimensions, the space-time metric is written as follows,
\begin{equation}
ds^2 = g_{AB}\,dx^A\,dx^B = e^{2 A (y)} \, \eta_{\mu\nu}\,dx^{\mu}\,dx^{\nu} - d y^2, 
\label{001}
\end{equation}
where $\eta_{\mu\nu}\equiv\{+1,-1,-1,-1\}$, $\mu$ and $\nu$ run from $0$ to $3$, $y \equiv x_4$ is the infinite extra-dimension coordinate (varying from $-\infty$ to $\infty$) such that the normal to surfaces of constant $y$ lie orthogonal to the brane, $e^{2 A (y)}$ is the warp factor \footnote{For the purpose of our calculations, we have suppressed brane tension terms (tensionless brane). It can be assumed that  tension terms are absorbed by the metric (see, for instance Eqs.~(24) and (25) from Ref. \cite{Folomeev} and Refs. \cite{Bronnikov,Abdyrakhmanov} where the real scalar field Lagrangian is discussed in the context of thick brane solutions).}.

Considering the real scalar field action, Eq.~(\ref{000}), one can compute the stress-energy tensor
\begin{equation}
T_{AB}^{\chi} = \partial_A \chi\partial_B \chi + g_{AB}\,V(\chi) - \frac{1}{2}g_{AB}\,g^{MN} \partial_M \chi \partial_N \chi,
\label{002}
\end{equation}
which, supposing that both the scalar field and the warp factor dynamics depend only on the extra coordinate, $y$, leads to an explicit dependence of the energy density in terms of the field, $\chi$, and of its first derivative, $d\chi/dy$, as
\begin{equation}
T_{00}^{\chi} (y) =\left[\frac{1}{2}\left(\frac{d\chi}{dy}\right)^2 + V(\chi)\right]\,  e^{2A(y)}.
\label{003}
\end{equation}

With the same constraints on $\chi$ about the dependence on $y$, the equations of motion arising from the above action are
\begin{equation}
\frac{d^2\chi}{dy^2} + 4 \frac{d A}{dy} \frac{d \chi}{dy} - \frac{d}{d\chi}V(\chi) = 0,
\label{004}
\end{equation}
by varying the action with respect to the scalar field, $\chi$, and
\begin{equation}
\frac{3}{2}\frac{d^2 A}{dy^2} = -  \left(\frac{d\chi}{dy}\right)^2,
\label{005}
\end{equation}
by varying the action with respect to the metric, or equivalently to $A$, which can be manipulated to yield 
\begin{equation}
3 \left(\frac{d A}{dy}\right)^2 = \frac{1}{2} \left(\frac{d\chi}{dy}\right)^2 - V(\chi).
\label{006}
\end{equation}
after integrating over $y$.

The potential for the real scalar field can be written in terms of a {\em superpotential}, $w$, in a specialized form as 
\begin{equation}
V(\chi) = \frac{1}{8}\left(\frac{dw}{d\chi}\right)^2 - \frac{1}{3} w^2,
\label{007}
\end{equation}
which has been often discussed in the context of thick brane solutions with a single scalar field \cite{DeWolfe,Gremm,Erlich,Afonso,Sasakura}.
It has the advantage of simplifying the above equations through first-order equations
\begin{equation}
\frac{d\chi}{dy} = \frac{1}{2}\frac{d w}{d\chi},
\label{008}
\end{equation}
and
\begin{equation}
\frac{d A}{dy} = - \frac{1}{3} w,
\label{009}
\end{equation}
for which analytical solutions can be immediately obtained through simple integrations.
for which analytical solutions can be immediately obtained through simple integrations.
In particular, it was first discussed in the context of supergravity on domain walls \cite{Rey} and its corresponding generalization to non-supersymmetric domain walls in various dimensions \cite{DeWolfe,Skenderis}.
Another method through which one endows the scalar field dependence on the extra-dimension and obtains the metric function and the potential through the field equations have been discussed \cite{Kobayashi,Slatyer} (see also Ref.~\cite{Folomeev} and references therein).

From Eq.~(\ref{007}) follows the energy density expressed as
\begin{equation}
T_{00}^{\chi}(y) = \left[\frac{1}{4}\left(\frac{dw}{d\chi}\right)^2 - \frac{1}{3} w^2\right]\,  e^{2A(y)}\label{0031}.
\end{equation}
As will be discussed next, an analogous first-order formulation for tachyon fields can be carried out.

\subsection*{Born-Infeld tachyon fields}

The action for a tachyon field, $\varphi$, coupled to $5$-dimensional gravity is given by Eq.~(\ref{000B}), in the geometry described by Eq.~(\ref{001}).
The tachyon field and the warp factor depend only on $y$ and allow for computing the stress-energy tensor
\begin{equation}
T_{AB}^{\varphi} (y)=
U(\varphi)\, \partial_A \varphi\partial_B \varphi \frac{1}{\sqrt{1 - g^{MN} \partial_M \varphi \partial_N \varphi}} + g_{AB}\,U(\varphi)\, \sqrt{1 - g^{MN} \partial_M \varphi \partial_N \varphi},
\label{002B}
\end{equation}
from which one also obtains the energy density as
\begin{equation}
T_{00}^{\varphi}(y) =  e^{2A(y)} \,U(\varphi)\,\sqrt{1 + \left(\frac{d\varphi}{dy}\right)^2}.
\label{003B}
\end{equation}

Under the same assumptions about the $\varphi$ dependence on $y$, the equations of motion can be obtained by varying the action with respect to the scalar field, $\varphi$, as
\begin{equation}
\frac{d^2\varphi}{dy^2} + \left[1 + \left(\frac{d\varphi}{dy}\right)^2\right]
\left( 4 \frac{d A}{dy} \frac{d \varphi}{dy} -\frac{1}{U(\varphi)} \frac{d}{d\varphi}U(\varphi)\right) = 0,
\label{004B}
\end{equation}
and by varying the action with respect to the metric (or $A(y)$) as
\begin{equation}
\frac{3}{2}\frac{d^2 A}{dy^2} = \left(\frac{d \varphi}{dy}\right)^2\,
\frac{U(\varphi)}{\sqrt{1 + \left(\frac{d\varphi}{dy}\right)^2}},
\label{005B}
\end{equation}
which can be manipulated in order to give
\begin{equation}
3 \left(\frac{d A}{dy}\right)^2 = 
 - \frac{U(\varphi)}{\sqrt{1 + \left(\frac{d\varphi}{dy}\right)^2}}.
\label{006B}
\end{equation}

Thus, once a potential for the tachyon field can be written as, for instance,
\begin{equation}
U(\varphi) = -\frac{3}{\upsilon^2}\sqrt{1 + \frac{1}{4} \left(\frac{d\upsilon}{d\varphi}\right)^2 },
\label{007B}
\end{equation}
where another {\em superpotential}, $\upsilon$, is introduced, one obtains the first-order equations,
\begin{equation}
\frac{d\varphi}{dy} = \frac{1}{2}\frac{d\upsilon}{d\varphi},
\label{008B}
\end{equation}
and
\begin{equation}
\frac{d A}{dy} = - \frac{1}{\upsilon},
\label{009B}
\end{equation}
such that the energy density Eq.~(\ref{003B}) can be written as
\begin{equation}
T_{00}^{\varphi}(y) = - \frac{3}{\upsilon^2}\left[1 + \frac{1}{4}\left(\frac{d\upsilon}{d\varphi}\right)^2\right]\,e^{2A(y)}.
\label{0031B}
\end{equation}

The energy densities Eq.~(\ref{003}) and Eq.~(\ref{003B}) can be shown to be the same through the relationship between the superpotentials, $w$ and $\upsilon$, 
\begin{equation}
\upsilon\left(\varphi(y)\right) \, w\left(\chi(y)\right) = 3.
\label{010}
\end{equation}
This relationship results into an equivalence between the Born-Infeld tachyon and the real scalar field dynamics.
Indeed, from Eqs.~(\ref{009}) and (\ref{009B}), one obtains
\begin{equation}
\left(\frac{d \chi}{dy} \right)^2 = - 3 \left(\frac{d A}{dy}\right)^2\, \left(\frac{d \varphi}{dy} \right)^2,
\label{012}
\end{equation}
through which, from Eqs.~(\ref{0031}) and (\ref{0031B}), and after some straightforward mathematical manipulations, it follows that
\begin{eqnarray}
T_{00}^{\varphi}(y) 
&=& - \frac{3}{\upsilon^2}\left[1 + \frac{1}{4}\left(\frac{d\upsilon}{d\varphi}\right)^2\right]\,e^{2A(y)}\nonumber\\
&=& - 3\left(\frac{d A}{dy}\right)^2 \,\left[1 + \left(\frac{d\varphi}{dy}\right)^2\right]\, e^{2A(y)}\nonumber\\
&=& \left[\left(\frac{d\chi}{dy}\right)^2 - 3\left(\frac{d A}{dy}\right)^2\right]\, e^{2A(y)}\nonumber\\
&=&\left[\frac{1}{4}\left(\frac{dw}{d\chi}\right)^2 - \frac{1}{3} w^2\right]\,  e^{2A(y)}\nonumber\\
&=&T_{00}^{\chi}(y).
\label{013}
\end{eqnarray}
In fact, the above result can be extended to the entire stress-energy tensor:
\begin{eqnarray}
T_{ij}^{\varphi}(y) 
&=& g_{ij}\, e^{- 2A(y)}\, T_{00}^{\varphi}(y) \nonumber\\
&=& g_{ij}\, e^{- 2A(y)}\, T_{00}^{\chi}(y) \nonumber\\
&=&T_{ij}^{\chi}(y),
\label{013BBB}
\end{eqnarray}
and $T_{0j} = T_{i0} = 0$ for $i,\,j = 1,\,...,\,4$.

To illustrate such an equivalence between two distinct models for brane structures, let us consider two examples, $I$ and $II$, for which the warp factor, $A(y)$, and the energy density, $T_{00}(y)$, can be analytically computed.

In terms of a real scalar, model $I$ is introduced through a sine-Gordon inspired superpotential given by
\begin{equation}
w^I(\chi) = \frac{2}{\sqrt{2} a} \sin{\left(\sqrt{\frac{2}{3}} \chi\right)},
\label{014}
\end{equation}
which reproduces the results previously obtained in Ref. \cite{Gremm}.
The model $II$ consists in a deformed $\lambda \chi^4$ theory with the superpotential
\begin{equation}
w^{II}(\chi) = \frac{3\sqrt{3}}{a} \left(1 - \frac{\chi^2}{9}\right).
\label{0141}
\end{equation}
In both cases, $a$ is an arbitrary parameter to fix the thickness of the brane described by the warp factor, $e^{2A(y)}$.
As expected, through Eq.~(\ref{008}), the superpotentials $w^I$ and $w^{II}$ lead to the respective solutions for $\chi(y)$,
\begin{equation}
\chi^{I}(y) = \sqrt{6} \arctan{\left[\tanh\left(\frac{y}{2\sqrt{2}a}\right)\right]},
\label{015}
\end{equation}
and
\begin{equation}
\chi^{II}(y) = 
3 \, \mbox{sech}{\left(\frac{\sqrt{3}y}{2a}\right)},
\label{0151}
\end{equation}
where, for convenience, we have just considered the positive solutions \footnote{In Eq.~(\ref{015}) there could let be explicit a constant of integration that amounts to letting $y \rightarrow y + C$, corresponding to the position of the brane in the extra dimension, for which we have set $C = 0$.}.

\begin{figure}
\centerline{\psfig{file= 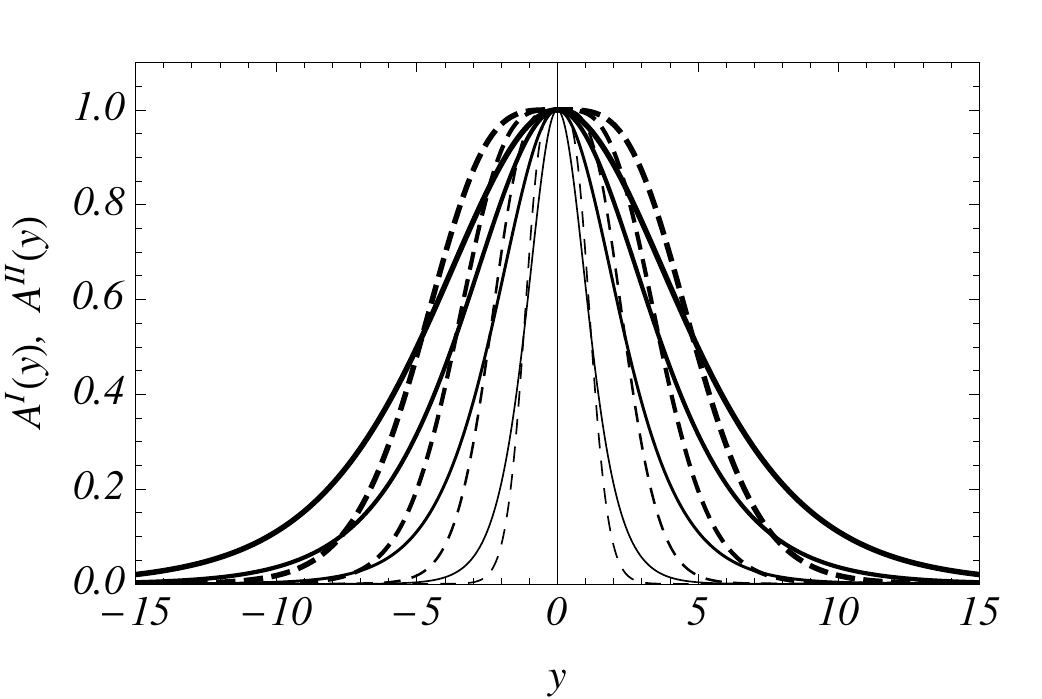, width= 10 cm}}
\caption{Warp factor, $e^{2A(y)}$, for models I (solid lines) and II (dashed lines) for a parameter $a$ running from $1$ (thinest line) to $4$ (thickest line), implying an increasing thickness.}
\label{brane01}
\end{figure}

\begin{figure}
\centerline{\psfig{file= 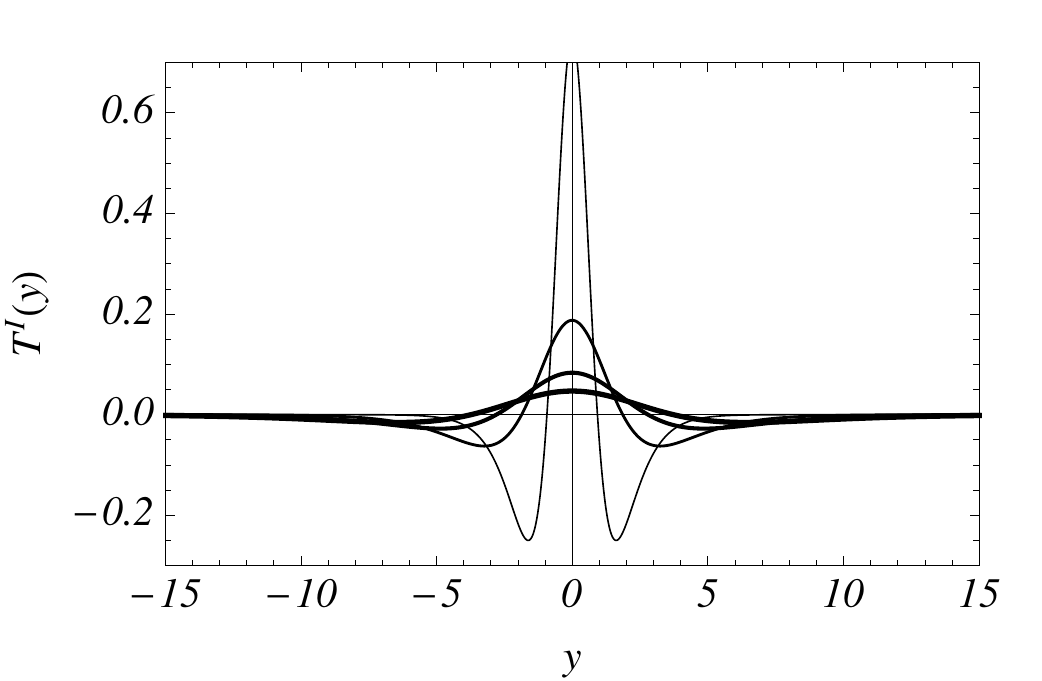, width= 10 cm}}
\centerline{\psfig{file= 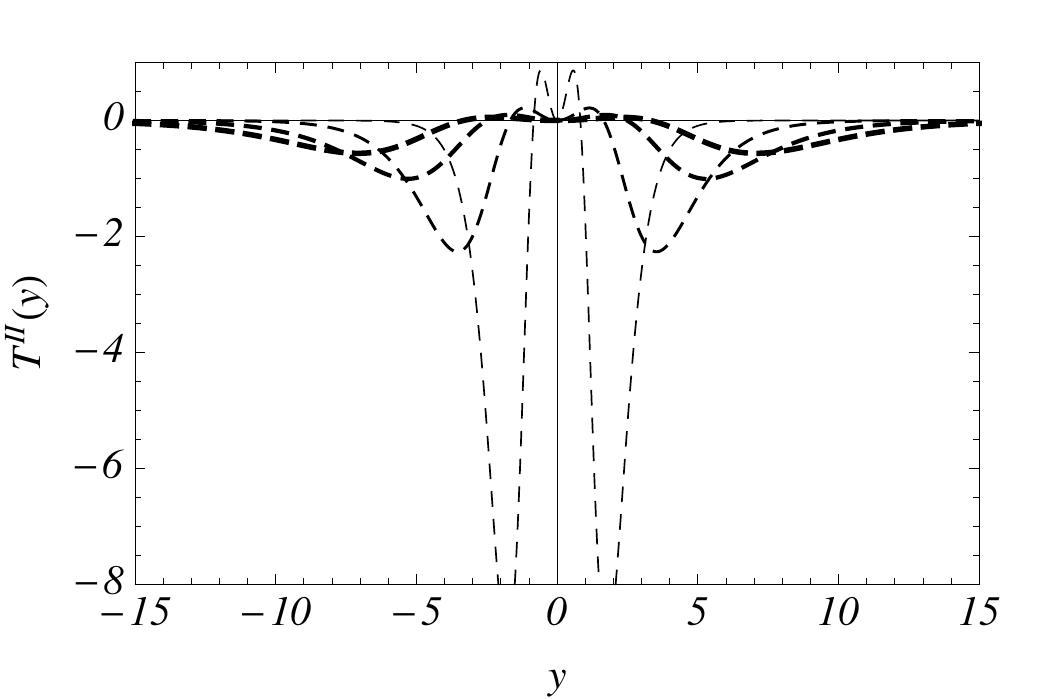, width= 10 cm}}
\caption{Energy density, $T_{00}(y)$, for models I (solid lines) and II (dashed lines) with parameter $a$ running from $1$ (thinest line) to $4$ (thickest line).}
\label{brane02}
\end{figure}

The corresponding expressions for the warp factor are obtained from Eq.~(\ref{008}) and are respectively,
\begin{equation}
A^{I}(y) = - \ln{\left[\cosh\left(\frac{y}{\sqrt{2}a}\right)\right]},
\label{016}
\end{equation}
\begin{equation}
A^{II}(y) = \tanh{\left(\frac{\sqrt{3}y}{2a}\right)}^2 - 2 \ln{\left[\cosh\left(\frac{\sqrt{3}y}{2a}\right)\right]},
\label{0161}
\end{equation}
where we have adopted the normalization criterium that sets $A(0) = 0$.
The solutions for $A^{I}$ and $A^{II}$ are depicted in Fig.~\ref{brane01}.
One can observe that both models $I$ and $II$ give rise to thick branes with the corresponding localized energy densities (c. f. Eq.~(\ref{0031})) given respectively by
\begin{equation}
T^{I}_{00}(y) = \frac{3}{4a^2}\mbox{sech}\left(\frac{y}{\sqrt{2}a}\right)^2
\left[\mbox{sech}\left(\frac{y}{\sqrt{2}a}\right)^2 - 2 \tanh\left(\frac{y}{\sqrt{2}a}\right)^2\right],
\label{020}
\end{equation}
and
\begin{equation}
T^{II}_{00}(y) = \frac{9}{a^2}\, e^{2\tanh\left(\frac{\sqrt{3}y}{2a}\right)^2}
\left[\mbox{sech}\left(\frac{\sqrt{3}y}{2a}\right)\tanh\left(\frac{\sqrt{3}y}{2a}\right)\right]
^2 \left[\frac{3}{4}\mbox{sech}\left(\frac{\sqrt{3}y}{2a}\right)^4 - \tanh\left(\frac{\sqrt{3}y}{2a}\right)^4\right],
\label{0201}
\end{equation}
which are depicted in Fig.~\ref{brane02}.

\begin{figure}
\centerline{\psfig{file= 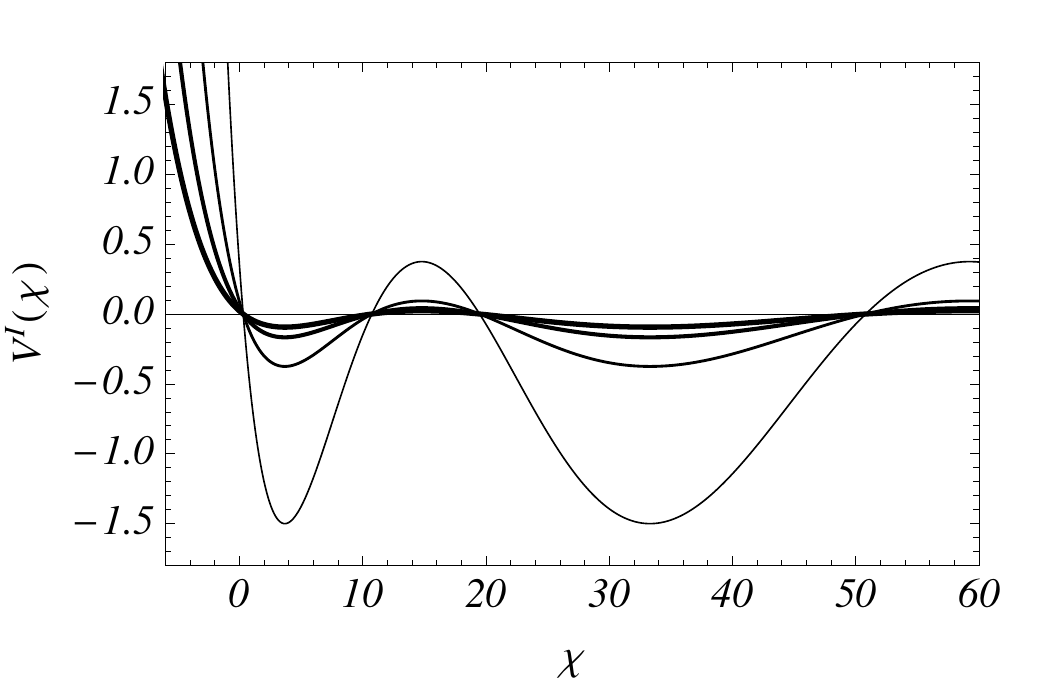, width= 10 cm}}
\centerline{\psfig{file= 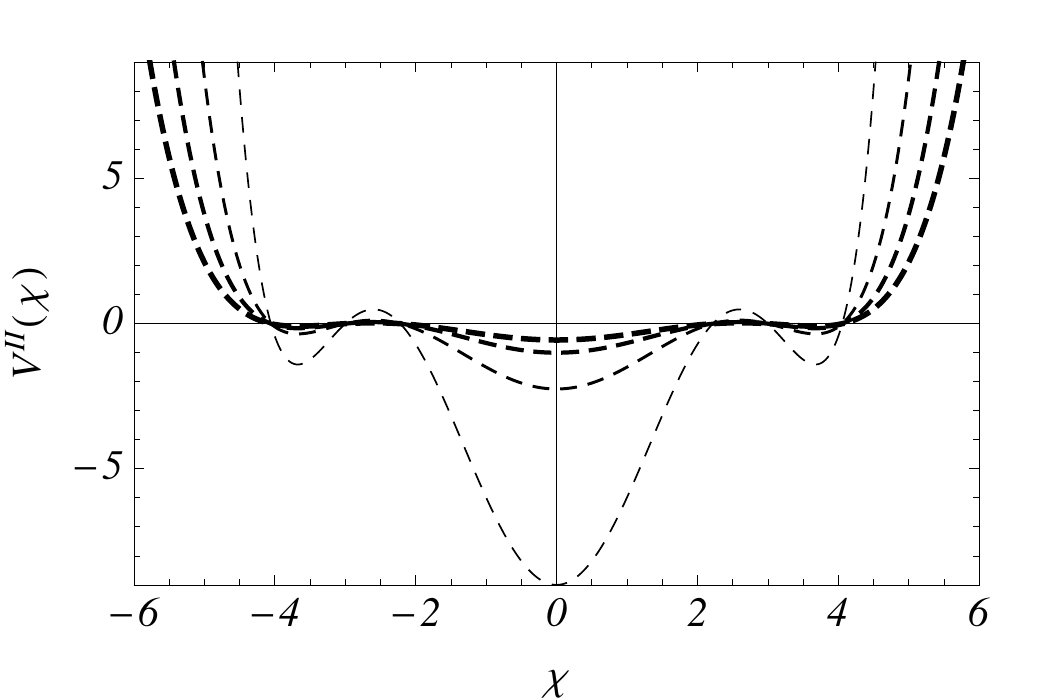, width= 10 cm}}
\caption{Potentials $V^{I,II}(\chi)$ for models I (solid lines) and II (dashed lines) with parameter $a$ running from $1$ (thinest line) to $4$ (thickest line).}
\label{brane04}
\end{figure}

The two Born-Infeld models are obtained via the corresponding superpotentials, $\upsilon^{I,II}(y)$, through the constraint Eq.~(\ref{010}).
They satisfy the following first-order equations for $\varphi$,
\begin{equation}
\frac{d\varphi^{I}}{dy} = \pm \frac{i}{\sqrt{2}} \frac{1}{\sinh{(y/\sqrt{2}a)}},
\label{017}
\end{equation}
and
\begin{equation}
\frac{d\varphi^{II}}{dy} = \pm \frac{\sqrt{3} i}{2} \frac{\cosh{(\sqrt{3}y/2a)}}{\left[\sinh{(\sqrt{3}y/2a)}\right]^2},
\label{0171}
\end{equation}
whose corresponding solutions are respectively:
\begin{equation}
\varphi^{I}(y) = \pm i \, a \,\ln{\left[\tanh{\left(\frac{y}{2\sqrt{2}a}\right)}\right]},
\label{018}
\end{equation}
\begin{equation}
\varphi^{II}(y) = \mp i \, a \,\mbox{csch}{\left(\frac{\sqrt{3}y}{2a}\right)}.
\label{0181}
\end{equation}

Finally, the corresponding Born-Infeld tachyon potentials are given respectively by
\begin{equation}
U^{I}(\varphi) = -\frac{3}{2\sqrt{2}a^2} \sec{\left(\frac{\varphi}{a}\right)}
\left[2 \sec{\left(\frac{\varphi}{a}\right)}^2 + \tan{\left(\frac{\varphi}{a}\right)}^2\right]^{\frac{1}{2}},
\label{022}
\end{equation}
and
\begin{equation}
U^{II}(\varphi) = -\frac{9}{a^2}
\left[1 + \frac{3}{2}\frac{\varphi^2}{a^2}\left(1- \frac{\varphi^2}{a^2}\right)\right]^{\frac{1}{2}}
\left(1- \frac{\varphi^2}{a^2}\right)^{-3},
\label{0221}
\end{equation}
which correspond to the effective real scalar field potentials,
\begin{equation}
V^{I}(\chi) = \frac{3}{8 a^2}\left[1 - 5 \sin{\left(\sqrt{\frac{2}{3}}\chi\right)}^2\right],
\label{023}
\end{equation}
and
\begin{equation}
V^{II}(\chi) = -\frac{1}{a^2}\left(1 - \frac{\chi^2}{9}\right)\left(9 - \frac{19\chi^2}{8} + \frac{\chi^4}{9}\right).
\label{0231}
\end{equation}
Potentials $V^I(\chi)$ and $V^{II}(\chi)$ suggest the possibility of SSB as it is depicted in Fig.~\ref{brane04}.
However, despite giving rise to the same brane structures, the potentials of the Born-Infeld models, $U^I(\varphi)$ and $U^{II}(\varphi)$, do not hint any SSB.
They correspond to a {\em plateu}-shaped potential with unstable dynamics, with a {\em plateu}-width proportional to $a$.

In order to relate the above results with some features of tachyonic models \cite{Zhang,Pal,Folomeev} one could replace the constraint Eq.~(\ref{010}) by
\begin{equation}
\upsilon\left(\varphi(y)\right) \, w\left(\chi(y)\right) = - 3,
\nonumber
\label{010BB}
\end{equation}
which corresponds to change the relative sign between the superpotentials, $w(y)$ and $\upsilon(y)$.
This would also give rise to $AdS$ domain walls with unlimited energy densities for the tachyon fields.
As an example, we consider the case of some tachyonic models described through the correspondence with model $I$ (c. f. Eq.~(\ref{023})), where $w\left(\chi(y)\right)$ is replaced by $- w\left(\chi(y)\right)$ in order to match the constraint Eq.~(\ref{010BB}).
This leads to 
\begin{eqnarray}
U^I(\psi(y)) &=& \frac{3}{2 a^2} \mbox{sech}{\left(\frac{y}{\sqrt{2}a}\right)}^2
\left[\sinh{\left(\frac{y}{\sqrt{2}a}\right)}^4 + \frac{1}{2}\sinh{\left(\frac{y}{\sqrt{2}a}\right)}^2\right]^{\frac{1}{2}},
\end{eqnarray}
which corresponds to the solutions of Refs.~\cite{Zhang,Pal,Folomeev}, if it is assumed a constraint between the $5$-dimensional cosmological constant, $\Lambda_5$, and the Hubble expansion rate, $H$, namely $\Lambda_5 = 6 H$.

Once the correspondence between tachyon and real scalar field solutions has been established, a second issue to point out concerns the difficulty in obtaining analytical expressions from the integration of the superpotentials $\upsilon(\varphi)$, i. e. sometimes the integrals the would result into the explicit dependence of $\varphi$ on $y$ could have no analytical representation.
To illustrate this point, one could study some deformed topological solutions departing, for instance, from superpotentials like 
\begin{equation}
w^{III}(\chi) = \frac{2}{a} \arctan\left[\sinh(\chi)\right],
\end{equation}
or
\begin{equation}
w^{IV}(\chi) = \frac{1}{4a}\left[\chi\left(5 - 2 \chi^2\right)\sqrt{1 - \chi^2} 
+ 3 \arctan\left(\frac{\chi}{\sqrt{1 - \chi^2} }\right)\right],
\end{equation}
which have been considered in discussions about deformed and multi- defects \cite{Bazeia,Bernardini,Ahmed}.
They give rise to the following solutions for $\chi(y)$:
\begin{equation}
\chi^{III}(y) = \mbox{arcsinh}\left(\frac{y}{a}\right),
\end{equation}
\begin{equation}
\chi^{IV}(y) = \frac{y}{\sqrt{a^2 + y^2}},
\end{equation}
and, from Eq.~(\ref{009}), the warp factors can be computed,
\begin{equation}
A^{III}(y) = \frac{1}{3}\left[\ln\left(1 + \frac{y^2}{a^2}\right) - 2\frac{y}{a} \arctan\left(\frac{y^2}{a^2}\right)\right],
\end{equation}
\begin{equation}
A^{IV}(y) = -\frac{1}{12}\left[\frac{y^2}{\sqrt{a^2 + y^2}} + 3\frac{y}{a} \arctan\left(\frac{y}{a}\right)\right],
\end{equation}
corresponding to thick brane solutions which induce the stability of the subjacent geometry.
However, for cases $III$ and $IV$, the correspondence with tachyonic solutions cannot be established analytically.

Finally, for models with gravity coupled to scalars, one cannot discuss the quantum fluctuations of the metric around the background without including scalar perturbations as well.
The treatment of scalar fluctuations is a rather complicated issue since it does not allow for an analytical treatment.
Otherwise, the gravity wave sector of the metric fluctuations decouples from the scalars \cite{DeWolfe,Gremm,Brane} so that it can be treated analytically.

In this case, the issue of stability of the metric fluctuations can be verified by assuming that a perturbed metric can be written as
\begin{equation}
ds^2 = e^{2 A (y)} \,\left( \eta_{\mu\nu} + h_{\mu\nu}\right)\,dx^{\mu}\,dx^{\nu} - d y^2, 
\label{001F}
\end{equation}
where we extracted a factor $e^{2 A (y)}$ from the fluctuation term to simplify subsequent equations, with $h_{\mu\nu} \equiv  h_{\mu\nu} (x,y)$ in the form of transverse and traceless tensor perturbations, for which one has the equation of motion \cite{DeWolfe},
\begin{equation}
\left(\frac{d^2~ }{dy^2} + 4\frac{d A}{dy}\frac{d~ }{dy} - e^{-2A(y)}\square\right) h_{\mu\nu}(x,y) = 0,
\label{001G}
\end{equation}
for linearized gravity decoupled to the scalar field, where $\square \equiv \partial_\mu\partial^{\mu}$.
As one can notice, the equation of motion for the metric perturbations corresponds to the Einstein equation of perturbations through $\delta G_{\mu\nu} = \delta T_{\mu\nu}$, where $G_{\mu\nu}$ is the Einstein tensor.
Therefore, assuming that the warp factors for the actions (\ref{000B}) and (\ref{000}) are the same (c. f. Eq.~(\ref{010})), the same corresponding stress-energy tensors from Eqs.~(\ref{013}) and (\ref{013BBB}) guarantee that Eq.~(\ref{001G}) remains the same for $\chi(y)$ and $\varphi(y)$.
In this particular scenario, Eq.~(\ref{001G}) for $h_{\mu\nu}$ can be written in terms of $A(y)$ for both actions, (\ref{000B}) and (\ref{000}).

Assuming a solution of the form
\begin{equation}
h_{\mu\nu} (x,z) = e^{i\, k.x}\, e^{-(3 A(z)/2)}\, H_{\mu\nu}(z),
\end{equation}
with $dz = e^{-A(y)}\,dy$, and dropping the index from $H_{\mu\nu}$, one can transform Eq.~(\ref{001G}) into a Schr\"odinger-like equation,
\begin{equation}
-H^{\prime\prime}(z) + V_{(QM)}(z)\, H(z) = k^2\, H(z),
\end{equation}
such that the localized zero-mode solutions ($k = 0$) for $4$-dimensional gravitational waves can be obtained through the study of the potential
\begin{equation}
V_{(QM)}(z) = \frac{3}{2} A^{\prime\prime}(z) + \frac{9}{4} A^{\prime2}(z),
\end{equation}
where the primes denote derivative with respect to $z$.
It is possible to state that all the above solutions, from models $I$ to $IV$, induce stability of the underlying geometry of the problem if $V_{(QM)}$ corresponds to volcano-type potentials induced by thick warp factors.
Indeed, it can be verified that the zero-modes of models $I$ to $IV$ correspond to the ground-state of $V_{(QM)}$, which gives rise to stable scenarios (i. e. $k^2 > 0$).
In this case, specifically for the graviton sector, one should expect no tachyonic modes such that no tachyonic condensation takes place.

To conclude, we can say that we have found, through first-order equations of motion, a relationship between Born-Infeld tachyon and real scalar solutions corresponding to an identical energy density.
In what concerns stability, the obtained solutions are all stable under metric perturbations provided that the effective volcano-type potential in the associated Schr\"odinger-like problem leads to normalizable ground state zero-mode wave functions.\\

{\em Acknowledgments - The work of A. E. B. is supported by the Brazilian Agencies FAPESP (grant 12/03561-0) and CNPq (grant 300233/2010-8). The work of O. B. is partially supported by the Portuguese Funda\c{c}\~ao para Ci\^encia e Tecnologia (FCT) by the project PTDC/FIS/11132/2009.}

\end{document}